\PassOptionsToPackage{english}{babel}
\documentclass[%
 reprint,
superscriptaddress,
 amsmath,amssymb,
 aps,abbrv,
prl,
floatfix,
]{revtex4-1}

\usepackage[english]{babel}

\usepackage{graphicx}
\usepackage{dcolumn}
\usepackage{bm}
\usepackage{hyperref}
\usepackage[mathlines]{lineno}

\renewcommand{\vec}[1]{\mathbf{#1}}

\begin{document}

\preprint{APS/123-QED}

\title{A Luneburg lens for spin waves}

\author{N. J. Whitehead}
\affiliation{Department of Physics \& Astronomy, University of Exeter, Stocker Road, Exeter, EX4 4QL, UK.}
\author{S. A. R. Horsley}
\affiliation{Department of Physics \& Astronomy, University of Exeter, Stocker Road, Exeter, EX4 4QL, UK.}
\author{T. G. Philbin}
\affiliation{Department of Physics \& Astronomy, University of Exeter, Stocker Road, Exeter, EX4 4QL, UK.}
\author{V. V. Kruglyak}
\email{V.V.Kruglyak@exeter.ac.uk}
\affiliation{Department of Physics \& Astronomy, University of Exeter, Stocker Road, Exeter, EX4 4QL, UK.}

\date{\today}

\begin{abstract}
We report on the theory of a Luneburg lens for forward-volume magnetostatic spin waves, and verify its operation via micromagnetic modelling. The lens converts a plane wave to a point source (and vice versa) by a designed graded index, realised here by either modulating the thickness or the saturation magnetization in a circular region. We find that the lens enhances the wave amplitude by 5 times at the lens focus, and 47\% of the incident energy arrives in the focus region. Furthermore, small deviations in the profile can still result in good focusing, if the lens index is graded smoothly. 

\end{abstract}

\maketitle



It is often useful to manipulate a wave as it travels through a material, and this can be achieved by designing a suitable graded refractive index. This is a well-established field in optics \cite{moore_gradient-index_1980}, and the techniques have also been applied to other areas of wave physics. 

In magnonics \cite{kruglyak_magnonics_2010, serga_yig_2010}, the study of spin waves, the theme of `graded index magnonics' \cite{davies_graded-index_2015} has been gaining interest recently as we begin to explore the many parameters of magnetic materials that can be manipulated to confine \cite{bayer_spin-wave_2003, tartakovskaya_spin_2016}, direct \cite{davies_towards_2015-1, gruszecki_spin-wave_2018} or generate \cite{davies_generation_2015-1, davies_generation_2016, whitehead_theory_2017} spin waves. 

In graded index optics, one well-known profile is the Luneburg lens \cite{luneburg_mathematical_1964}, a rotationally-symmetric refractive index profile designed to focus a plane wave to a point, or conversely, to convert a point source to a plane wave. This profile has been studied in many other areas of wave physics \cite{falco_luneburg_2011, zentgraf_plasmonic_2011, dockrey_thin_2013, kim_sound_2014}, due to its applications for use with antennas. As such, the Luneburg lens may have an important role in future wave-based computing circuitry, to launch plane waves from an antenna, or increase the amplitude of incoming plane waves to be read by the same antenna. To read/launch a plane wave from/to a different direction, one only needs to move the antenna to the corresponding point on the edge of the lens, without having to reconfigure the lens.

In this work, we demonstrate theoretically how a Luneburg lens for spin waves may be realised in a magnetic thin film. 

\begin{figure}
\centering
\includegraphics[width=\linewidth]{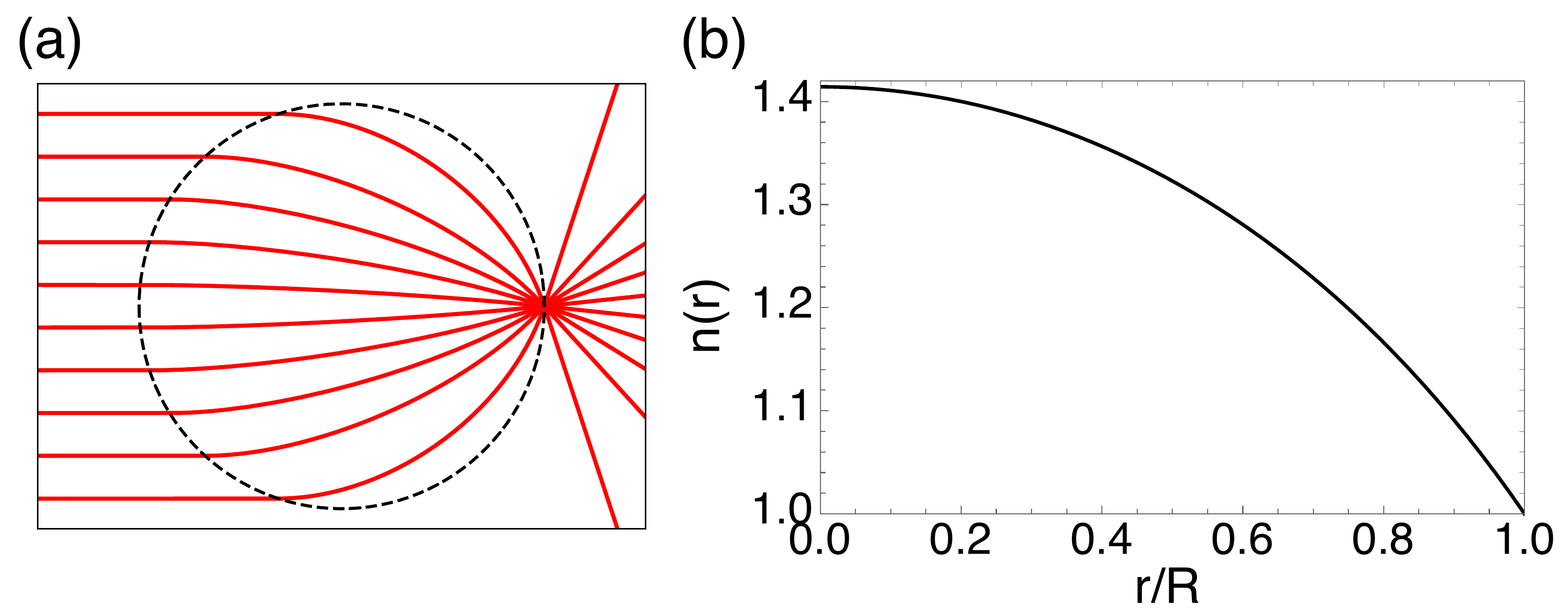}
\caption{(Color online) (a) The Luneburg lens, outlined by the black dashed line, focuses rays (red lines) to a diffraction-limited spot on the opposite edge of the lens. (b) Refractive index profile for the lens.}
\label{fig:fig1_profile}
\end{figure}

The refractive index profile $n(r)$ for a Luneburg lens is given by
\begin{equation}
\label{eq:luneburg}
n(r)=\sqrt{2-(r/R)^2},
\end{equation}
where $r$ is the radial coordinate and $R$ is the radius of the lens. This profile, along with the ideal operation of the lens, is shown in Fig. \ref{fig:fig1_profile}. 

For light propagation in an isotropic material, the refractive index is given by $n=c/v$, with $c$ and $v$ being the speed of light in vacuum and the medium, respectively. In a dispersive medium however, there is both a phase index, which relates to the phase velocity of light as before, and a group index, related to the group velocity. When these indices are the same, i.e. the dispersion relation for frequency $\omega$ as a function of wave-number $k$ is approximately linear and there is no band gap in the spectrum, the graded index has the same spatial profile for different frequencies.

For spin waves, the ferromagnetic medium is always dispersive. Furthermore, the dispersion relation $\omega(\vec{k})$ may depend upon the mutual orientation of the wave-vector and magnetization. The `magnonic index' is therefore both anisotropic and frequency-dependent, however we will only work with spin waves in a narrow frequency range. To make a spin wave Luneburg lens, we thus need to ensure that
\begin{equation}
\label{eq:luneburg_sw}
n(r) = \frac{k(r)}{k_\text{ref}}=\sqrt{2-(r/R)^2},
\end{equation}
where $k(r)$ is the required wave-number at radial coordinate $r$, and $k_\text{ref}$ is the reference wave-number outside of the lens. To achieve this, we choose to work with forward-volume magnetostatic spin waves, propagating in the plane of a perpendicularly magnetized thin ferromagnetic film. The waves have the following isotropic dispersion relation \cite{stancil_spin_2009-1}:
\begin{align}
\label{eq:disp}
k =\frac{1}{s} \left[\text{arctan} \left(\frac{1}{\sqrt{-(1+\kappa)}} \right) \right] \frac{2}{\sqrt{-(1+\kappa)}},
\end{align}
where
\begin{align}
\label{eq:def}
\kappa &= \frac{\Omega_H}{\Omega_H^2-\Omega^2}, \quad \Omega = \frac{\omega}{4\pi\gamma M_S}, \quad \Omega_H = \frac{H_i}{4\pi M_S},
\end{align}
$M_S$ is the saturation magnetization, $H_i=H_0-4\pi M_S$ is the internal magnetic field, $H_0$ is the applied external field, and $s$ is the film thickness. 

\begin{figure}
\centering
\includegraphics[width=\linewidth]{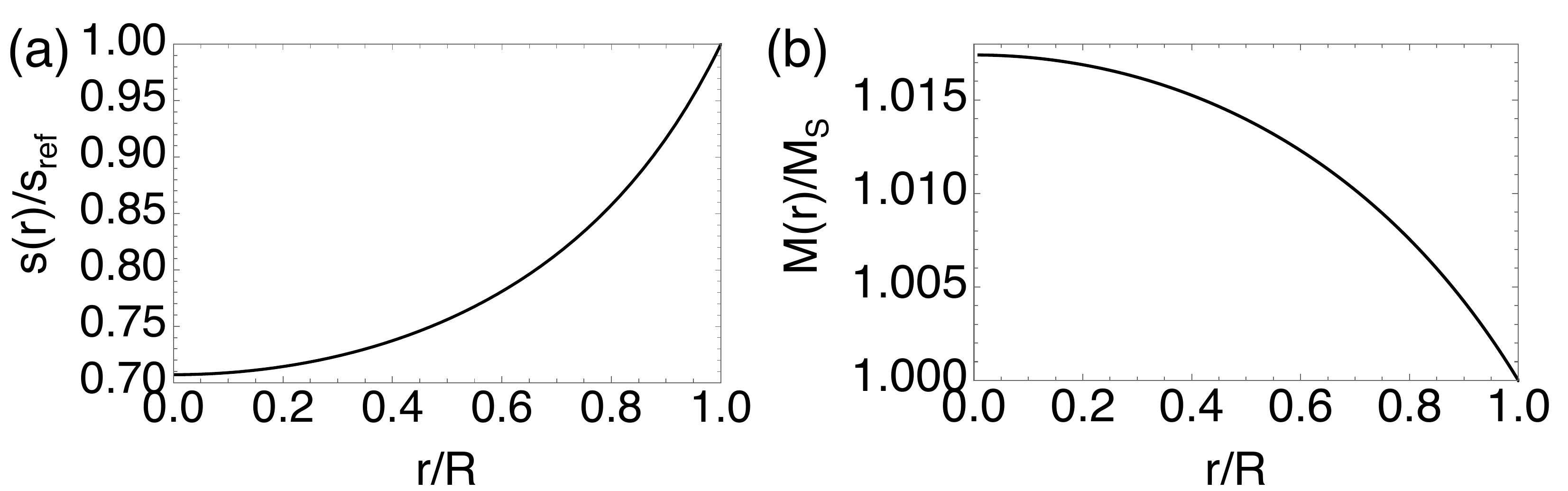}
\caption{The (a) thickness profile and (b) magnetization profile required to make a Luneburg lens.}
\label{fig:fig2_sm}
\end{figure}

There are three parameters in equations \eqref{eq:disp} and \eqref{eq:def} that we can manipulate to vary the wave-number, and hence the index: $s$, $M_S$ and $H_0$. Interestingly, if we neglect any change of the internal field with thickness, and keep all other quantities the same, there is a very simple relation between the index and the film thickness outside the lens, $s_\text{ref}$, and inside the lens, $s(r)$,
\begin{equation}
\label{eq:s}
k(r)/k_\text{ref}= s_\text{ref}/s(r).
\end{equation}
The resulting profile of $s(r)$ is given in Figure \ref{fig:fig2_sm} (a). We can see that the centre of the profile is $1/\sqrt{2}$ times the original film thickness, or around 70\%. The large change in thickness required for the operation of the lens suggests that it will not be too sensitive to small thickness variations.

Although there is no simple relation between the index $k(r)/k_\text{ref}$ and the magnetization or external field, it is significantly easier to model changes in these quantities in micromagnetic simulations. So, to demonstrate the operation of a spin wave Luneburg lens, we choose to vary the saturation magnetization in space.  We determine the magnetization profile $M(r)$ required to produce the required lensing effect from the implicit relations \eqref{eq:disp} and \eqref{eq:def}, as shown in Fig. \ref{fig:fig2_sm} (b). There are two important features to notice. Firstly, $M(r)$ needs to increase in the center of the lens. Secondly, the required maximum change in the magnetization is just 1.7\%, which is small in comparison to the corresponding change in thickness required. This is due to the complicated dependence of $\kappa$ on $M_S$, in \eqref{eq:def}.

We now describe how the lens is designed and tested in micromagnetic simulations using MuMax3 \cite{vansteenkiste_design_2014-1} software. In the model, we create a 1 mm $\times$ 0.5 mm Yittrium-Iron-Garnet (YIG)-like film in the $x-y$ plane, with fixed thickness $s=2\mu$m and damping constant $\alpha = 1\times10^{-4}$. The saturation magnetization is $M_S= 140 $kA/m outside of the lens and varies inside the lens as shown in Fig. \ref{fig:fig2_sm} (b). The cell size is 0.5 $\mu$m in the $x-y$ film plane, with periodic boundary conditions in both in-plane directions. We choose to have 1 cell across the thickness, which is much smaller than the wavelength $\lambda$ of the studied spin waves. We apply an out-of-plane bias field $H_\text{ex}=200$ mT, and then apply a burst of microwave magnetic field with central frequency $f = 1$ GHz (corresponding to $\lambda =$33.9 $\mu$m), bandwidth of $0.1f$ and amplitude of $0.1$ mT, directed along the $x$-axis, which in turn generates a spin wave packet. 

 The Luneburg lens is only designed to be effective in the geometrical optics approximation\cite{leonhardt_chapter_2009}, i.e., when the wavelength is much smaller than the lens size, which determines the radius of the lens, $R$. From comparison with other studies \cite{mattheakis_luneburg_2012, de_pineda_broadband_2017}, and to keep the simulation size reasonable, we use $R\approx6\lambda$. We then use \eqref{eq:s} to determine the radii of 255 concentric circular regions, between which the saturation magnetization changes in uniform steps to form the lens profile $M(r)$. 
\begin{figure}
\centering
\includegraphics[width=\linewidth]{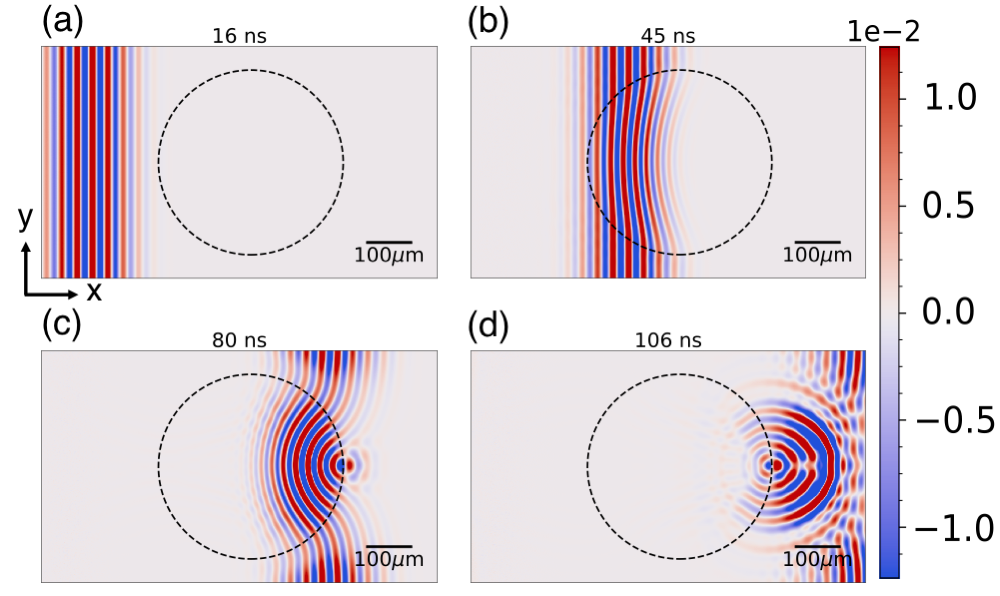}
\caption{(Color online) Snapshots of the $m_x$ component of the magnetisation as the wave packet moves through the Luneburg lens (dotted black line) at timesteps of (a) 16 ns, (b) 45 ns, (c) 80 ns and (d) 106 ns. Color scale is saturated to show the wave packet clearly.}
\label{fig:fig3_mx}
\end{figure}

\begin{figure}
\centering
\includegraphics[width=\linewidth]{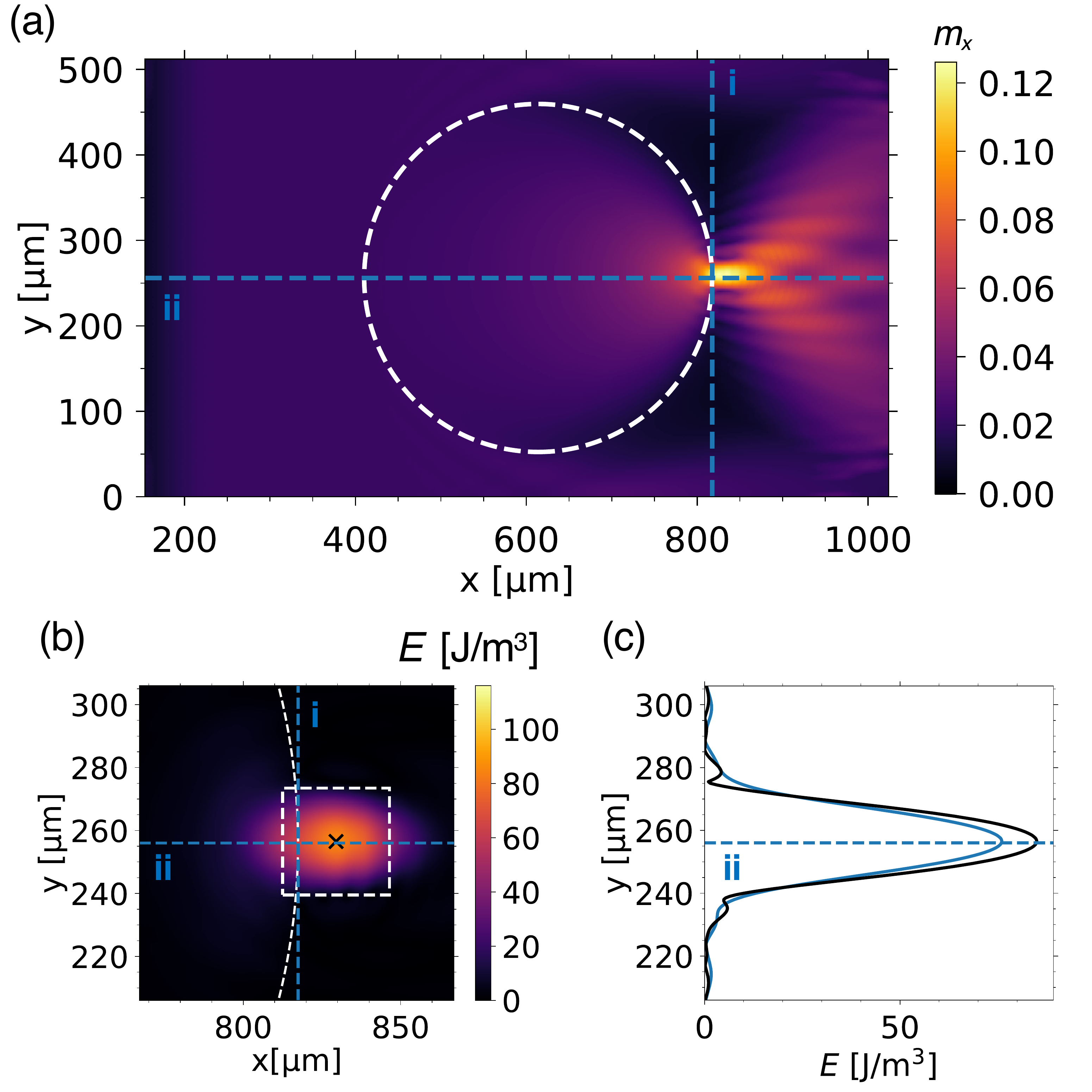}
\caption{(Color online) (a) Maximum amplitude of $m_x$ attained across the model over the entire duration of the simulation (omitting region before pulse starting point). (b) Energy density near focus region at the time of peak amplitude. The white square has side lengths of $\lambda$ and is centred on peak of actual focus spot (indicated with a black cross). Image (c) shows the energy density along the line i from (a) and (b) (blue line) and at the $x$ position of the actual focus (black line), at the times where the maximum amplitude occurs.}
\label{fig:fig4_amp}
\end{figure}

The snapshots of the normalized $x$-component of magnetization $m_x$ are shown in Fig. \ref{fig:fig3_mx} for different moments of time. In the Supplemental Material \footnote{See Supplemental Material for animations of Figures \ref{fig:fig3_mx}, \ref{fig:fig4_amp} (a) and \ref{fig:fig6_ps}, and an analysis of the negative wave-numbers produced in time.}, the corresponding video is also provided. The wavefronts behave as we would expect: the wavelength $\lambda$ decreases in the region of increased refractive index, and so curves the wavefronts towards the focus of the lens. In addition, the wavefronts are slowed in the central region, compared to the those outside. We see the effect of this after the wave has left the lens, where the focused energy appears to be re-emitted from the focal spot. Another point to note is that all of the frequencies in the wave packet appear to be focused equally well by the lens, despite the fact that it is only designed to work at the central frequency.

The degree of focusing is more clearly shown in Fig. \ref{fig:fig4_amp} (a), which gives the maximum amplitude of $m_x$ attained in each cell of the model, over the entire duration of the simulation. The corresponding video in the Supplemental material shows the amplitude of the wave moving through the lens in time. We can clearly identify that the largest amplitude is attained in the focus region of the lens. In Fig. \ref{fig:fig4_amp} (b) we show the energy density near the focus region, at the time that the maximum amplitude attained in the entire simulation is achieved. The energy is mostly concentrated around $\pm \lambda/2$ of the ideal focus. However, the focus peak is actually shifted along $x$ from the ideal position, similar to the observation reported in Ref. \onlinecite{rozenfeld_electromagnetic_1976}. Increasing the size of the lens with respect to $\lambda$ should bring the focal spot closer to the edge of the lens. In Fig. \ref{fig:fig4_amp} (c), we plot the energy density along the line i shown in (a) and (b), and also along $y$ for the $x$ position of the actual focus peak. The full width at half maximum (FWHM) of the peak at the actual focus is around 23 $\mu$m or 0.67$\lambda$, which is reasonable for a diffraction-limited lens. At the actual focus and the ideal focus, the peak amplitudes of $m_x$ are 5 and 4.7 times larger than the unfocused amplitude, respectively. This enhancement of the wave amplitude may be useful when reading an incoming plane wave using an antenna.
  
To investigate the efficiency of the lens, we can compare how much of the wave packet energy reaches the focal region of the lens. We do this by summing the energy density in each cell in a rectangular region before the lens, and in the $\lambda\times\lambda$ region centred on the actual focus peak shown in Fig. \ref{fig:fig4_amp} (b), and comparing the values. The rectangular region has an $x$ extent of 300 $\mu$m, which completely encompasses the width of the pulse, and $y$ extent of $2r_\text{lens}$, horizontally aligned with the lens centre. We find that 46\% of the energy is measured in the focal region at the time of peak amplitude. This is the proportion of the incoming wave that could be measured by an antenna centred at the focal peak. In the Supplemental Material, we consider a different method to quantify the lens efficiency. We compare the Fourier amplitudes for the positive and negative wave-numbers that result from the wave's interaction with the lens, and find that no more than 13\% of the wave is reflected overall.
\begin{figure}
\centering
\includegraphics[width=\linewidth]{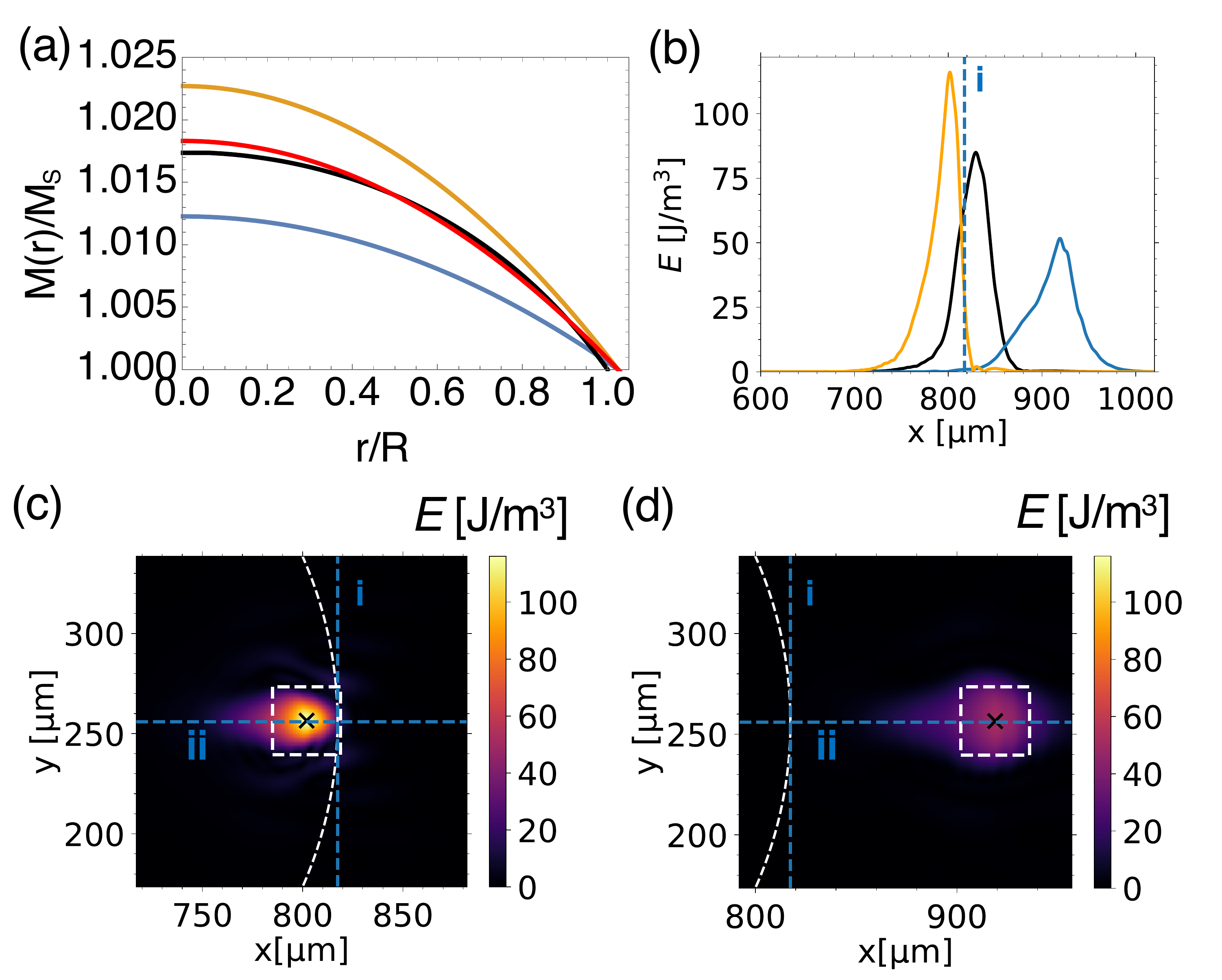}
\caption{(Color online) (a) Comparing the Luneburg profile (black) with a parabola, with either a 5\% (red), or $\pm30$\% (orange, blue respectively) error in $M(0)$. (b) Depth of focus plot showing energy density along $x$ at the $y$ position of the actual focus, for the profiles with corresponding colors in (a), with the ideal focus $x$ position (line i) from (c) and (d) shown. Plots of energy density across $x$ and $y$ for (c) $+30\%$ and (d) $-30\%$ $M(0)$ error are shown near the ideal Luneburg lens focus (intersection of lines i and ii, all other features as per Fig. \ref{fig:fig4_amp} (b)). }
\label{fig:fig5_para}
\end{figure}

Finally, we examine how sensitive the results are to deviations in the Luneburg magnetization profile. We must ensure that $M(R)=M_S$, so we just need to alter the value of the magnetization in the centre of the lens, $M(0)$. We fit the profile to the form $M(r) = 1 + \alpha \, \{a - [b \, (r/R)^2] - 1\}$, where $\alpha$ is a parameter that changes $M(0)$, and $a$ and $b$ are constants determined by the fit. The profiles for the best fit, and poorer fits with errors of $\pm30$\% are given in Fig. \ref{fig:fig5_para} (a). Notice that the radius of the lens has to change slightly to accommodate the parabolic fit, and for the index to be equal to 1 at the lens edge, so $R_\text{fit} =1.025R$. 

The best fit, with $\alpha=1$, has an $M(0)$ error of 5\% and yet still produces an almost identical lensing effect to the actual Luneburg profile; the spin wave amplitude at the actual focal spot is still 5 times larger than that outside of the lens. We do not show this result here --- rather, to aid comparison with the previous results, we compare the results of the $\pm30\%$ error profiles to the actual Luneburg profile in Fig. \ref{fig:fig5_para}. If $M(0)$ is increased, the lensing effect is increased as we would expect, creating a narrower focus and increasing the amplitude of $m_x$ at the focus peak to 5.7 times for an $M(0)$ error of +30\%. If $M(0)$ is reduced below the ideal value, the $m_x$ amplitude reduces more substantially to 3.8 times for an $M(0)$ error of -30\%. In \ref{fig:fig5_para} (b) we show the energy density along the depth of focus for each profile, and we find that the +30\% error, Luneburg, and -30\% error profiles have FWHMs of 0.8$\lambda$, 1.1$\lambda$ and 1.3$\lambda$ respectively.

We now compare the energy of the wave packet before entering the lens with the energy in the $\lambda\times\lambda$ focus regions shown in Fig. \ref{fig:fig5_para} (c) and (d) \footnote{Note that in Fig. \ref{fig:fig5_para} (c), (d), and Fig. \ref{fig:fig4_amp} (b), the colour scale is set to the maximum value attained in Fig. \ref{fig:fig5_para} (c), for ease of comparison}. We find that 49\% of the original energy arrives in the focus region for the +30\% error profile, compared with 37\% for the -30\% error profile. Recall that the Luneburg profile received 46\% of the incident energy at the focus region. So the +30\% error profile is certainly a tighter focus, with a larger amplitude and greater intensity in a smaller region. However, the focus position is within the lens in this case, which is a disadvantage. 

Overall, it seems that the curvature of the lens profile can deviate from the ideal Luneburg profile and still produce a reasonable lensing effect. However, only the actual Luneburg profile can both successfully focus a plane wave to a spot and convert a point source to a plane wave, as we show in Fig. \ref{fig:fig6_ps}. To create these images (and their associated videos in the Supplemental Material) we have used the same parameters as previously, except extended the size of the model (to reduce interference due to the periodic boundary conditions) and introduced a continuous point source with a Gaussian profile. The point source is either located at the ideal focus on the edge of the lens, or at the actual focus determined from the focussing study, and we compare the Luneburg lens to the $\pm 30$ \% $M(0)$ error profiles as before. We can see that the Luneburg profile is the most successful in creating a plane wave, albeit with some interference due to reflections within the lens. The $+30$ \% profile focuses the outgoing wavefronts too much for both source positions, and the -30\% profile focuses the outgoing wavefronts too little.  Positioning the source at the actual focus in the latter case is a good improvement, but the wavefronts are still not completely parallel to each other, and the wave amplitude is still lower than in the other two cases.

\begin{figure}
\centering
\includegraphics[width=0.9\linewidth]{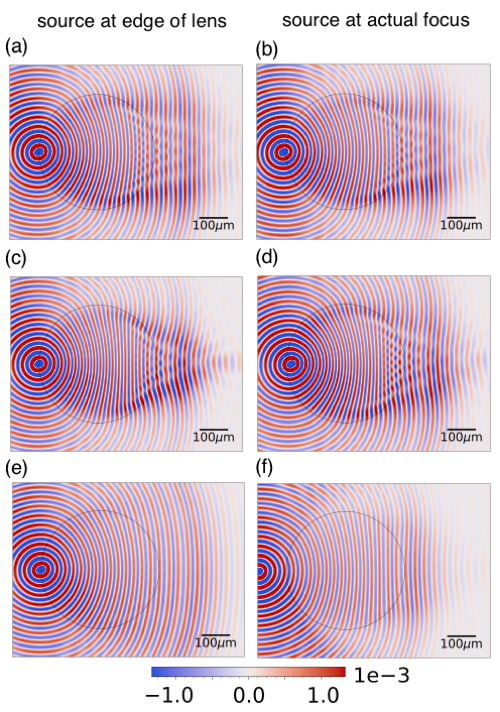}
\caption{Snapshots of $m_x$ at 82ns when a Gaussian point source is positioned near (a)-(b) the Luneburg lens, (c)-(d) the +30\% error profile, and (e)-(f) the -30\% error profile. The point source is centred on the lens edge for (a), (c) and (e), and on the actual focus position for (b), (d) and (f). As per the corresponding Supplemental animations, the color scale has been set to the same value for all plots, with scale at the bottom (saturated for clarity).}
\label{fig:fig6_ps}
\end{figure}

To conclude, we have demonstrated how to form a Luneburg lens in a magnetic material with perpendicular magnetization. There is a simple relation between the refractive index and the magnetic film thickness, with the centre of the Luneburg profile having a thickness of $1/\sqrt{2}$ times the original film thickness. We also demonstrate via micromagnetic modeling how to create a Luneburg lens by varying the saturation magnetization in the lens region, although in this case the change in magnetization required is only 1.7\% of the bulk value. The lens works effectively, increasing the amplitude of the wave by 5 times at the focus. As long as the magnetization (or any other parameter) is graded smoothly, the lens is reasonably insensitive to variations from the ideal Luneburg profile. The latter is still the optimal profile for focusing a plane wave to a point and converting a point source to a plane wave, as the source / detector could be positioned at the lens edge in both cases. 

\begin{acknowledgments}
We would like to thank F. B. Mushenok for helping to create the pulsed excitation in the micromagnetic modelling, and C. A. Vincent for technical support throughout. This research has received funding from the Engineering and Physical Sciences Research Council (EPSRC) of the United Kingdom, via the EPSRC Centre for Doctoral Training in Metamaterials (Grant No. EP/L015331/1), and the European Union's Horizon 2020 research and innovation program under Marie Sk\l{}odowska-Curie Grant Agreement No. 644348 (MagIC). SARH would like to thank the Royal Society and TATA for financial support.
\end{acknowledgments}


%

\pagebreak

\clearpage

\pagebreak

\setcounter{equation}{0}
\setcounter{figure}{0}
\setcounter{page}{1} 
\renewcommand{\thefigure}{S\arabic{figure}} 
\renewcommand{\theHfigure}{Supplement.\thefigure} 

\onecolumngrid 

\section{A Luneburg lens for spin waves: supplementary material}

\subsection{List of Supplementary Animations and their Captions}

Note that in Animations 3-5, the color scale has been set to 0.1 of the maximum and minimum values of $m_x$ recorded in Animation 4 (b), to ensure the wavefronts are clearly visible and that each are comparable.

Animation 1 - Time-dependence of $m_x$ as the pulse moves through the Luneburg lens. This is the animation corresponding to Fig. 3 in the main text, and in both cases the color scale has been saturated at 0.1 of the maximum recorded value to ensure the wavefronts are clearly visible. \\

Animation 2 - Time-dependence of the $m_x$ amplitude, as the pulse moves through the Luneburg lens. This is the animation corresponding to Fig. 4 (a) in the main text. Note that the scale has been set between 0 and the maximum recorded amplitude in this case. \\

Animation 3 (a) and (b) - Time-dependence of $m_x$ passing through the Luneburg lens, when the Gaussian point source is positioned at (a) the ideal focus (edge of lens) and (b) the actual focus. \\

Animation 4 (a) and (b) - Time-dependence of $m_x$ passing through the parabolic magnetization profile, with a +30\% error in $M(0)$, when the Gaussian point source is positioned at (a) the edge of the lens and (b) the actual focus. \\

Animation 5 (a) and (b) - Time-dependence of $m_x$ passing through the parabolic magnetization profile, with a -30\% error in $M(0)$, when the Gaussian point source is positioned at (a) the edge of the lens and (b) the actual focus. \\

Animation 6 - Time-dependence of $k_x - k_y$ Fourier amplitude. The scale has been set between 0 and the maximum recorded Fourier amplitude in this case.

\subsection{Lens Efficiency: $kx-ky$ Analysis}

An alternative way to quantify the reflection loss of the Luneburg lens is to consider the spin wave scattering in the reciprocal space. To do this, we Fourier transform the wave field in space, and compare the Fourier amplitudes of positive and negative $k_x$ components of the wave vector. To account for the wave phase, we save the data from the simulations at a time-step $\Delta t= \pi/(2\omega)$, with $\omega=2\pi f$, and construct a complex wave field. The real and imaginary components of this field at each time-step are given by the wave fields at adjacent time-steps, $\pi/2$ out of phase. 

\begin{figure}
\centering
\includegraphics[width=\linewidth]{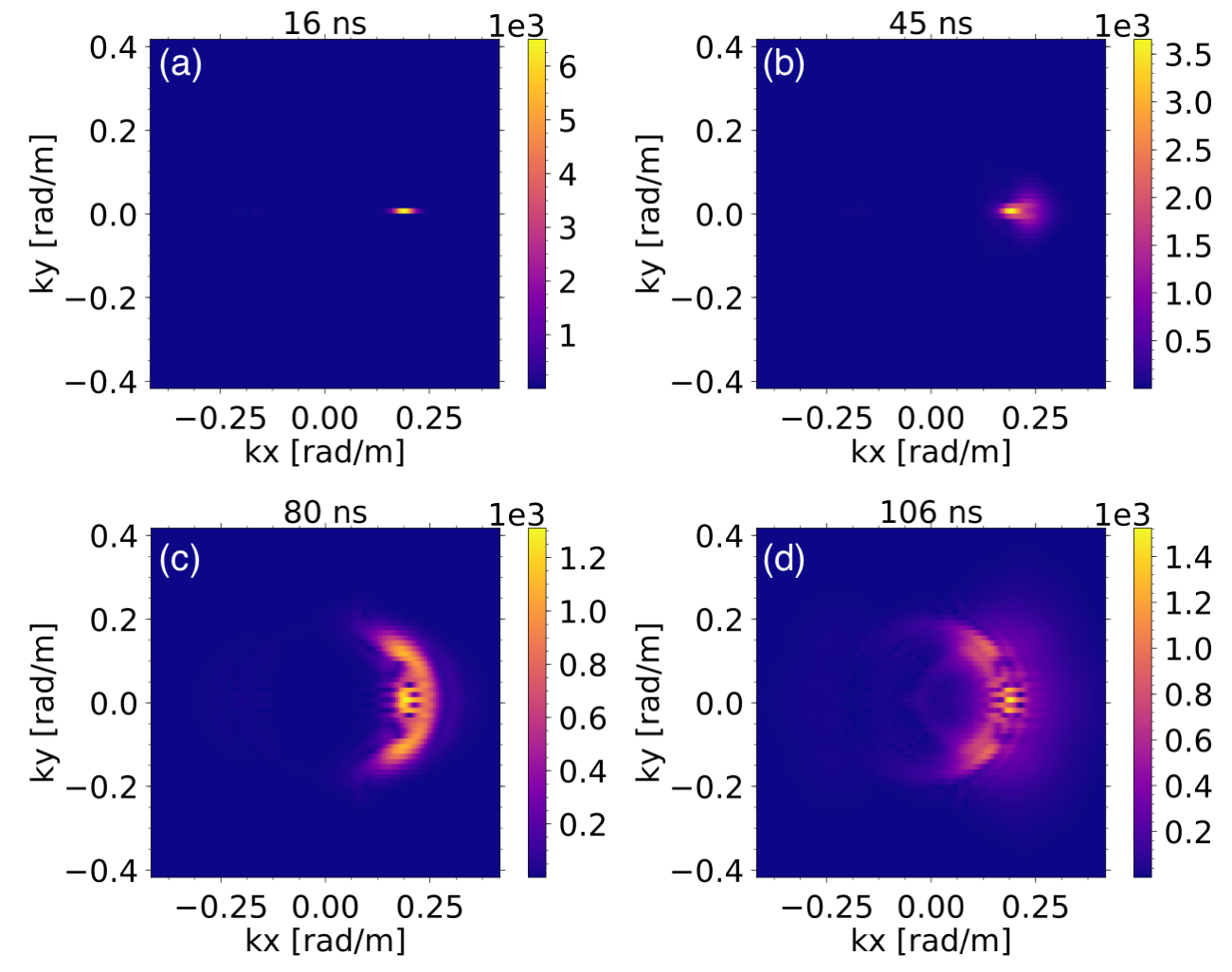}
\caption{Fourier amplitudes in $k_x-k_y$ space at (a) 16ns, (b) 45ns, (c) 80ns and (d) 106ns, which correspond to the same snapshots as in Fig. 3 in the main text. Note that the colour scale differs in each image for clarity, but each are in arbitrary (yet proportional) units.}
\label{fig:kxky_supp}
\end{figure}

\begin{figure}
\centering
\includegraphics[width=\linewidth]{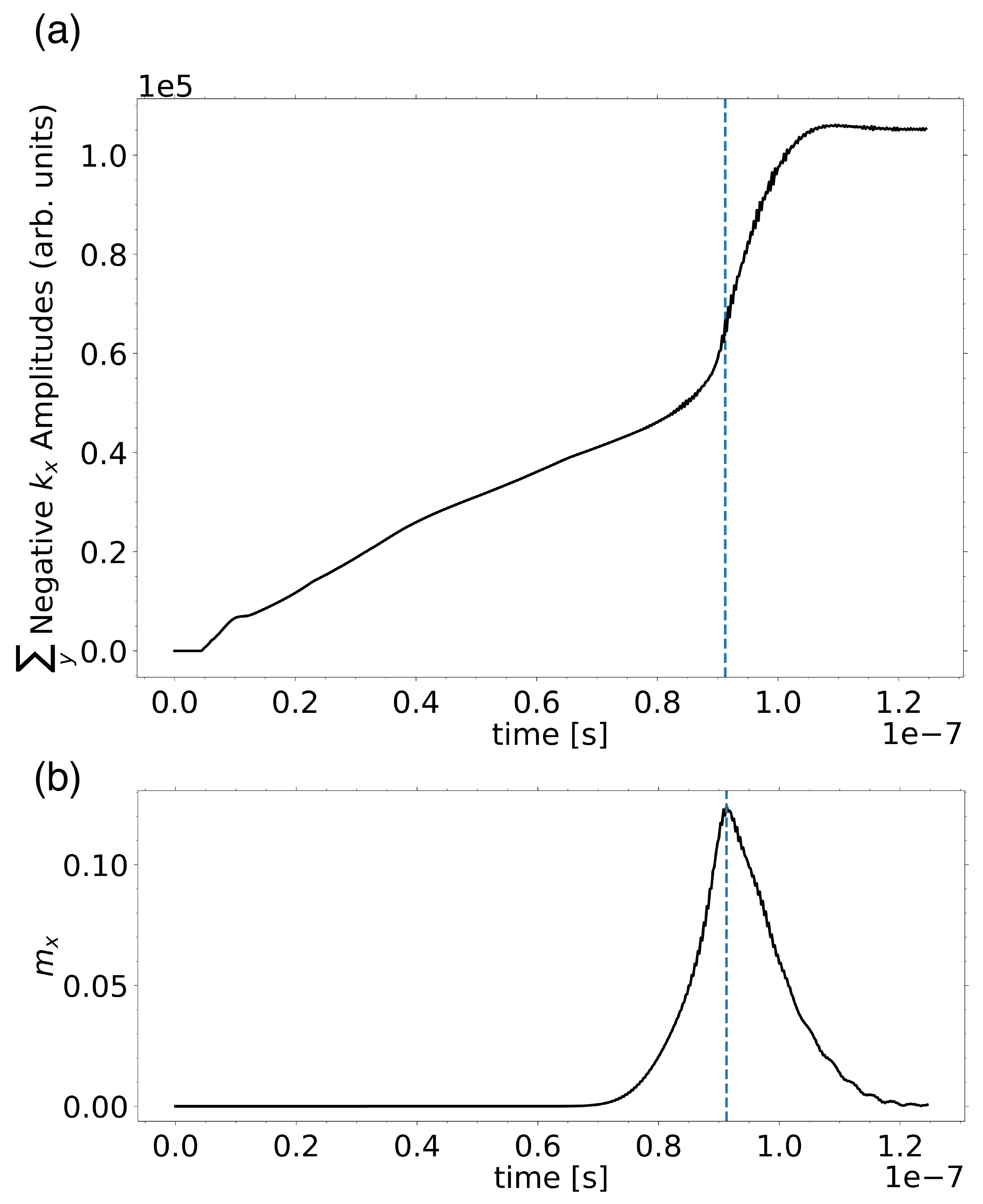}
\caption{(a) Accumulation of the negative $k_x$ Fourier amplitude (summed over all of $k_y$) over time. Blue dashed line indicates the time at which the peak of the pulse encounters the actual focus, as shown in (b), where $m_x$ is plotted at the centre of the actual focus spot in time.}
\label{fig:kxtime_supp}
\end{figure}

If we sum the absolute values of the Fourier amplitudes across all of $k_y$ and compare the amplitudes in positive $k_x$ to negative $k_x$, we find that 13\% of the wave is reflected in total. We would expect to see some reflection due to a perfect Luneburg lens anyway at each infinitesimal boundary where the refractive index changes. To confirm the origin of this reflection, we show how the Fourier amplitude changes in time as the wave moves through the lens in Fig. \ref{fig:kxky_supp}. These shapshots of the Fourier amplitude in the $k_x-k_y$ space are shown at times corresponding to the snapshots of $m_x$ in real space in Fig. 3 of the main text. The corresponding video is provided in the Supplemental Material. We can see that the Fourier amplitude of non-zero $k_y$ components emerges when the wavefronts become curved within the lens, and then spreads into a circle in (d) as the waves begin to emanate from the focus, as we would expect. The Fourier amplitude of negative $k_x$ components appears to be negligible until this point, suggesting that it is not the quality of the lens that reduces the proportion of energy transmitted through it. Rather, as the pulse exits the focus of the lens, the edges of the wavefronts seem to re-enter the lens as they spread out, and so travel in the negative $x$ direction, as can be seen in Animation 1. In Fig. \ref{fig:kxtime_supp}, we show how the negative $k_x$ components accumulate over time, by summing all of the negative $k_x$ values, over all $k_y$, at each time step. We see a reasonably steady increase of the negative $k_x$ amplitude until the peak of the pulse is nearing the centre of the actual focus. Around this time, the growth becomes steeper, which seems to confirm that this increase is not due to the pulse moving through the lens, but instead due to the wave spreading out and re-entering the lens after passing through the focus.

\clearpage

\end{document}